\documentclass[twocolumn,3p]{elsarticle}

\usepackage{graphicx,float, amsmath}
\usepackage[dvipsnames]{xcolor}
\usepackage[integrals]{wasysym} 
\usepackage{nicefrac,amssymb,amsmath} 

\biboptions{numbers,sort&compress} 
\begin{document}
 
\title{Phenomenological plasmon broadening and relation to the dispersion}

\author[]{Raphael Hobbiger\corref{cor1}}
\ead{raphael.hobbiger@jku.at}
\cortext[cor1]{Corresponding author}
\author[]{J\"urgen T. Drachta}
\author[]{Dominik Kreil}
\author{Helga M. B\"ohm}
\address{Institute for Theoretical Physics, Johannes Kepler University, A-4040 Linz, Austria}

\begin{abstract}
 Pragmatic ways of including lifetime broadening of collective modes in the electron liquid are critically compared.
 Special focus lies on the impact of the damping parameter onto the dispersion.
 It is quantitatively exemplified for the two-dimensional case, for both, the charge (`sheet'-)plasmon and the spin-density plasmon.
 The predicted deviations fall within the resolution limits of advanced
techniques.
\end{abstract}


\maketitle

\section{Introduction}

The study of plasmons, the collective oscillations of electrons, has a long and successful
history \cite{giuliani2005quantum,pines1999elementary}. Their coupling to light in nano--structures, known as 'plasmonics', holds high promise for revolutionary applications \cite{naik2013plasmonics},
the performance of actual devices being crucially limited by metallic losses \cite{Khurgin2014howto}.  
An undemanding inclusion of a plasmon's damping via a constant, irrespective of the losses' origin(s), is commonly achieved via a Drude type dielectric function \cite{pines1999elementary}.  
The response of charges, however, is non local, which is the more important the smaller the size of the nano--particles is.

The random phase approximation (RPA) \cite{pines1999elementary}, historically a milestone, provides a dielectric function explicitly depending on both, frequency \(\omega\) as well as wave vector \(q\).
Its failure to include finite lifetime effects
was treated early by Mermin \cite{MerminPhysRevB.1.2362}, his approach still being widely applied.
Recent examples in bulk systems include calculations of the electrons' inelastic mean free path \cite{NguyenTruong201479}, stopping power \cite{PhysRevE.90.053102}, and a generalization to spin wave damping \cite{HankiewiczVignale2008InhomogeneousGilbert}.
In layers, it has been employed, e.g.,
to obtain quasi--particle properties \cite{PhysRevB.53.9964} of the two--dimensional electron gas (2Deg), 
or when accounting for inter-band-excitation losses in graphene \cite{Jablan2009Plasmonicsingraphene}.

Major techniques for studying the charge response to external perturbations are scattering experiments \cite{StoegerPollach20081092} and, for long wavelengths, optical measurements.
They yield the same dispersion \textit{only\/} for an undamped plasmon; 
for realistic lifetimes slightly different results are obtained
\footnote{For the Drude case this was already noted in \cite{pines1999elementary}, p.207f.}.
This discrepancy in the plasmon {\it disper\-sion\/} \(\omega_{\mathrm{pl}}(q)\) increases with its linewidth (often broadening with \(q\)). 
Consequently, comparing theory and high-resolution experiments needs appropriate caution.

Forefront scattering data for dispersion and damping are known for metallic monolayers \cite{nagaoplasmonL,nagaoplasmonP,Rugeramigabo08} and semiconductor quantum wells \cite{Eriksson2000165,hirjibehedin2002evidence}.
Many 2Degs being rather dense \cite{PhysRevB.91.085204}, RPA predictions are sufficiently accurate, once damping effects are built in effectively.
The 2Deg, our prototype, shares the vanishing of \(\omega_{\mathrm{pl}}(q\!\to\!0)\) with graphene.
There, too, the plasmon was studied with optical as well as scattering methods 
\cite{FABZ11_DiracPlasmons,LBPS10_pldampbelow,DasSarma20121795,PhysRevB.88.085403,SKKK11_controlofpipls}
(plasmons in graphene, being thoroughly reviewed in \cite{Grigorenko2012Grapheneplasmonics,Abaj14_GraphenePlasmonics}).

Mermin's approach conserves the local electron number, invoking just a single additional constant \(\eta\) (the inverse lifetime \(\tau\)). 
Extensions further conserving local energy and momentum were developed (and applied to a two--component plasma) by R\"opke \cite{Roepke1998Dielectricfunction, PhysRevE.64.056410,PhysRevE.62.5648},
and, independently, by Atwal and Ashcroft \cite{PhysRevB.65.115109}.
These sophisticated theories yield intricate response functions with a wave vector dependent line--width, as also found in \cite{bohm2010dynamic}.
But in view of realistic materials an uncomplicated RPA extension incorporating plasmon lifetimes via a phenomenological (potentially \(q-\)dependent) parameter is preferential.
The purpose of this work is to critically compare such simple approaches with that of Mermin, and to study the resulting plasmon dispersions. 
Deviations from the classical plasma frequency, \(\omega_p\,\), turn out larger than expected.

After briefly discussing general aspects in Sec.\,\ref{sec: Theoretical Overview}, 
we present quantitative (zero temperature) results (both in RPA and beyond) in Sec.\,\ref{sec: Application to the 2Deg} for a 2Deg.
There, the vanishing of \(\omega_{\mathrm{pl}}(q\!\to\!0)\) implies a comparably
high relative width, also in case of rather weak damping.
Finally, we study the spin--density plasmon in the partially spin-polarized case in Sec.\,\ref{sec: Spin plasmon}.
For Fourier Transforms conventions are as in \cite{giuliani2005quantum}.

\section{Theoretical Overview}
\label{sec: Theoretical Overview}

A plasmon is conventionally obtained from the complex dielectric function 
\(\epsilon\! = \epsilon_{\scriptscriptstyle\mathrm I}\!+ i\, \epsilon_{\scriptscriptstyle\mathrm{II}}\)
via these, closely related but not exactly equal definitions:
\begin{subequations}
\begin{itemize}\addtolength\itemsep{-0.3em}
 \item as a maximum in the double differential scattering cross section,
   \begin{align}
    \text{Im}\,\frac{-1}{\epsilon\big(q,\omega_{\mathrm{pl}}^{\scriptscriptstyle\mathrm{(a)}}\big)}
    \>=\>\text{max} \;;
    \label{eq: plas_Imchi}
   \end{align}
 \item as the vanishing of the complex \(\epsilon\) for \hbox{complex} \(\omega\), determining reflection coefficients (purely oscillatory waves change to decaying ones)
   \begin{align}
     \epsilon\big(q,\,\omega_{_\mathrm I}\!+i\omega_{_\mathrm{II}}\big) \>=\> 0 \,,\;
     \omega_{\mathrm{pl}}^{\scriptscriptstyle\mathrm{(b)}} \equiv \omega_{_\mathrm I} \;;
    \label{eq: plas_compl}
   \end{align}
 \item 
 often approximated as the zero of \(\mathrm{Re}\,\epsilon\),
   \begin{align}
     \epsilon_{_\mathrm I}\!\big(q,\,\omega_{\mathrm{pl}}^{\scriptscriptstyle\mathrm{(c)}}\big) 
     \>=\> 0 \;;
    \label{eq: plas_Reeps}
   \end{align}
 (the approximation being justified if \(\omega_{\scriptscriptstyle\mathrm I}\!\gg\!\omega_{\scriptscriptstyle\mathrm{II}}\),
 and with \(\omega_{\mathrm{pl}}^{\scriptscriptstyle\mathrm{(c)}}\) always lower than \(\omega_{\mathrm{pl}}^{\scriptscriptstyle\mathrm{(a,b)}}\)
\cite{Mendlowitz:60}),
 \item or, for irrelevant phase shifts, as minimal magnitude of \(\epsilon\) (implying a maximal electric field), 
   \begin{align}
    \big|\epsilon\big(q,\omega_{\mathrm{pl}}^{\scriptscriptstyle\mathrm{(d)}}\big)\big| 
    \>=\> \text{min} \;.
    \label{eq: plas_magep}
   \end{align}
\end{itemize}  
\end{subequations}
The Drude model \cite{pines1999elementary} for charge carriers with a classical plasma frequency \(\omega_p\) reads
\(\epsilon_{\scriptscriptstyle\mathrm{Dru}}(q,\omega)= 1-\omega^2_p/\omega(\omega\!+\!i\eta)\).
Here, \(\omega_{\mathrm{pl}}^{\scriptscriptstyle\mathrm{(a)}}\) and \(\omega_{\mathrm{pl}}^{\scriptscriptstyle\mathrm{(b)}}\) differ \(<\!1\)\% for \(\eta\) even as large as \(\omega_p\), 
but conditions
(\ref{eq: plas_Reeps}--\ref{eq: plas_magep}) yield clearly distinct values, unless \(\eta\) is rather small.
The criterion appropriate to the setup must be chosen for cutting edge experimental resolutions.
Typical values reported \cite{nagaoplasmonL,Rugeramigabo08} are \(\sim\)10meV (roughly 10\%\(\,\omega_{\scriptscriptstyle{\rm pl}}\)), 
\(\sim\!25\!\ldots\!125\)meV \(\approx10\ldots50\,\)\%\,\(\omega_{\scriptscriptstyle{\rm pl}}\) \cite{LBPS10_pldampbelow} and \(\sim\,\)100meV \cite{RevModPhys.83.705}.
By definition, (\ref{eq: plas_Imchi}) yields a symmetric Lorentzian near \(\omega_{\mathrm{pl}}^{\scriptscriptstyle\mathrm{(a)}}\), whereas expansions around \(\omega_{\mathrm{pl}}^{\scriptscriptstyle\mathrm{(b-d)}}\) contain first order terms in the denominator, too:
\addtocounter{equation}{1}
\begin{align*}
 \quad\mathrm{Im}
 \frac{-1}{\epsilon(q,\,\omega)} &\approx\> 
 \left\{\begin{array}{lllll}
   \frac{\alpha}{\left((\omega\!-\!\omega_{\mathrm{pl}}^{\scriptscriptstyle\mathrm{(a)}}\right)^2
 \,+\, \gamma^2} 
 \hfill(\theequation)
  \vspace{0.2cm}\\
   \frac{\alpha}{\left(\omega\!-\!\omega^{\scriptscriptstyle\mathrm{(b-d)}}_{\mathrm{pl}}\right)^2 \,+\,
         \beta\left(\omega\!-\!\omega^{\scriptscriptstyle\mathrm{(b-d)}}_{\mathrm{pl}}\right)
\,+\, \gamma^2} \quad\;\;
 \end{array}\right.
\end{align*}

Clearly, the discrepancy in differently computed \(\omega_{\mathrm{pl}}-\)values depends on the specific \(\epsilon(q,\omega)\) used. Some common forms are given next.

The linear response of an electron liquid to external perturbations in RPA-type
approaches reads
\begin{align}
	\epsilon_{_{\mathrm{RPA}}}\!(q,\omega)\>=\>1-v(q)\,\chi^0(q,\omega) \;; 
 \label{eq:chiRPA}
\end{align}
here, \(v(q)\) denotes the Coulomb interaction, and \(\chi^0(q,\omega)\) the density-density response function of non-interacting fermions
 \cite{giuliani2005quantum}.
It shows the typi\-cal electron--hole (e/h) excitation band in the \((q,\omega)-\) plane.
An adiabatically turned on perturbation corresponds to \(\,\omega\to\omega + i0^{\scriptscriptstyle+}\).
This ensures causality, but yields an undamped plasmon.
An obvious idea to include damping is to use \(\,\chi^0(q,\widetilde\omega)\) with \(\,\widetilde\omega\!\equiv \omega\!+i\eta\) and inverse lifetime \(\eta\!\equiv1/\tau\),
\begin{align}
	\epsilon_{_{\mathrm{Lin}}}\!(q,\omega) \>\equiv\; 1-v(q)\>\chi^0(q,\,\widetilde\omega) \;.
 \label{eq:chiLin}
\end{align}
This also broadens the e/h band (as \(\text{Im}\,\chi^0(q,\,\widetilde\omega)\), at any \(q\), only vanishes when \(|\omega\tau|\gg1\)).
It catches the eye that Eq.\,(\ref{eq:chiLin}) alters the static response
\begin{align}
	\epsilon_{_{\mathrm{Lin}}}\!(q,0) &\;=\; 1-v(q)\,\chi^0(q,\,i\eta) 
 \\\nonumber
	&\;\xrightarrow[q\to0]{}\; 1 + \omega_p^2/\eta^2
\end{align}
violating \(\epsilon(q\!\to\!0,\,0) =\, 1\!+ v(q)\,N(E_{\scriptscriptstyle\mathrm{F}})\)
(the perfect screening sum rule, \(N(E_{\scriptscriptstyle\mathrm{F}})\) is the density of states at the Fermi energy). 
The correctness of this limit may be lesser in importance for plasmonic applications, which are far from static.

Mermin  \cite{MerminPhysRevB.1.2362} corrected the deficiency. He derived
\begin{subequations}
\label{eq:Merm}
\begin{align}
	\epsilon_{_{\mathrm{Me}}}\!(q,\omega) &\>\equiv\; 1-\frac{v(q)\,\chi^0(q,\,\widetilde\omega)}
        {1 + i\eta\,g_{_{\mathrm{Me}}}(q,\,\widetilde\omega)} \;; 
 \label{eq:chiMer}
 \\
	&\; g_{_{\mathrm{Me}}}\!(q,\,\widetilde\omega) \>\equiv\>
        \textstyle\frac1{\widetilde\omega}\big(\frac{\chi^0(q,\,\widetilde\omega)}{\chi^0(q,0)}-1\big) \;.
 \label{eq:gMer}
\end{align}
\end{subequations}
Albeit elegant, analytical calculations with Eq.\,(\ref{eq:Merm}) quickly get cumbersome, 
in particular when the relations are meant as matrix equations (\(2\!\times\!2\) for spin-dependent screening or electron-hole liquids, infinite matrices in crystals reciprocal lattice vectors).
Note that (\ref{eq:gMer}) does \textit{not\/} yield, as it should,  the classical plasmon for long wavelengths, 
\begin{equation}
   \omega_{\mathrm{pl}}^{\scriptscriptstyle\mathrm{(b)}}(q\!\to\!0)_{_{\mathrm{Me}}}
	\;\nrightarrow\; \omega_p \;,
   \label{eq: Merm wrong}
\end{equation}
neither in the bulk nor for the 2Deg \cite{PhysRevB.65.115109}
(there, also \(\omega_{\mathrm{pl}}^{\scriptscriptstyle\mathrm{(a)}}(q\!\to\!0)\)
 shows a mismatch with the \(\sqrt{q}-\) dependence, \emph{cf.} Fig.~1 below).

Comparing approaches with the structure of (\ref{eq:chiMer}) but
arbitrary \(\omega\) instead of \(\widetilde\omega\) is worthwhile,
\begin{align}
	\epsilon_g(q,\omega) &\>\equiv\; 1-\frac{v(q)\,\chi^0(q,\omega)}
        {1 + i\eta\,g(q,\omega)} \;.
 \label{eq:chig}
\end{align}
The elementary choice \(g_1\!\equiv\mathrm{sgn}(\omega_{_\mathrm I})\hbar/E_{\scriptscriptstyle\mathrm{F}}\) (\textit{i.e.\/}\ simply adding a constant to 
the RPA's susceptibility denominator, with a sign function for proper symmetry),
serves to enhance a long-lived plasmon's visi\-bility in graphical representations.
The dielectric function with \(g_{\scriptscriptstyle\mathrm{D}}\equiv 1/\omega\) reduces to the Drude model for \(\omega\!\gg \hbar q^2/2m\), (\(m\) is the effective electron mass).
For the optical conductivity \(\sigma\) this implies
\begin{align}
	\sigma(\omega) &\>=\; \frac{ne^2\tau/m}
        {\omega\,\big(g(0,\omega)-i\tau\big)} 
        \;\xrightarrow[g=g_{_{\mathrm{D}}}]{\,}\; \frac{ne^2\tau/m}{1-i\omega\tau} \;.
 \label{eq:sigm}
\end{align}

An interpolation between static RPA screening and the Drude case can be achieved by
\begin{align}
	g_{_{E_{\mathrm{F}}}}(\omega) \>\equiv\> \frac{\omega}{\omega^2 + E_{\mathrm{F}}^2/\hbar^2}\ .
 \label{eq:gEF}
\end{align}
The main deficiency of this ansatz is to violate the f-sum rule (due to additional poles at \(\hbar\omega\!= \pm i E_{\scriptscriptstyle\mathrm{F}}\)), 
\begin{equation}
	\int\limits_0^\infty\!\frac{d\omega}{\pi}\> \omega\>\mathrm{Im}\,\frac{-1}{\epsilon(q,\omega)} 
	\;=\> \frac{\omega_p^2}2 \;.
 \label{eq: w1SR}
\end{equation}
When the focus lies on plasmon properties (\emph{e.g.} the \(q-\)dependence), this can be acceptable:
No large frequency range (often inaccessible anyhow \cite{ColemanMBP2015}) needs to be measured for comparing peak positions and widths.

Both, \(\epsilon_{\scriptscriptstyle\mathrm{Me}}\) and \(\epsilon_{\scriptscriptstyle\mathrm{Lin}}\) fulfill Eq.\,(\ref{eq: w1SR}), for the latter occasionally reported otherwise \cite{PhysRevE.90.053102}.
The contour for the integration (\ref{eq: w1SR}) in the complex
\((\omega_{\scriptscriptstyle\mathrm{I}},\omega_{\scriptscriptstyle\mathrm{II}})-\)plane is taken
along the quarter circle enclosing the first quadrant.
For a purely real integration kernel on the
\(\omega_{\scriptscriptstyle\mathrm{II}}-\)axis only the arc contributes,
provided \(\mathrm{Im}\,\epsilon^{-1}\) is analytic in the upper half plane (as it should) \cite{giuliani2005quantum}.
The RPA response function obeys these conditions.
From its high frequency expansion,
\begin{align}
	\epsilon_{_\mathrm{RPA}\!}(q\!\to\!0,\, |\omega|\!\to\!\infty)
	\>\approx\> 1-{\omega_p^2}/{|\omega|^2} \;,
\label{eq: chi0expansion}
\end{align}
it follows that also \(\epsilon_{\scriptscriptstyle\mathrm{Lin}}\) fulfills the f-sum rule.

Note that the static structure factors \(S(q)\) obtained from the above dielectric functions via 
\begin{equation}
	\int\limits_0^\infty\!\frac{d(\hbar\omega)}{\pi}\>\mathrm{Im}\,\frac{-1}{\epsilon(q,\omega)} 
	\;=\> v(q)\,S(q) 
 \label{eq: w0SR}
\end{equation}
differ for each approach. 
For a meaningful comparison of scattering intensities, their  normalization by \(S(q\!\to\!0\)) appears advisable.

We point out that \(\eta\), entering as a parameter, does not necessarily coincide with the half width of the plasmon peak.
Only for \(\epsilon_{\scriptscriptstyle\mathrm{Lin}}\) the difference is small.
In particular, when the half width gets comparable with the distance between \(\hbar\omega_{\scriptscriptstyle\mathrm{pl}}(q)\) and the e/h-band, the agreement is worse.

\renewcommand{\arraystretch}{1.2}
\begin{table}[h]\begin{center}
\begin{tabular}{l|ccccccc}
        {\footnotesize crit./}  &{\footnotesize\(\mathrm{Im}\,\epsilon\)} 
       &{\footnotesize\(\!\!\epsilon(q,i\omega_{\scriptscriptstyle\mathrm{II}\!})\!\!\)\!} 
       &{\!\footnotesize caus.\!} &{\footnotesize\(\!\omega\!=\!0\!\)} &\!{\footnotesize f--SR}\!
       &\multicolumn{2}{c}
	{\footnotesize \(\;\;\omega_p^{\scriptscriptstyle\mathrm{class}} \!\!\!\)}
\vspace{-0.10cm}\\
 	{\footnotesize appr.}  &{\footnotesize\(>\!0\)}&{\footnotesize\(\!\in\mathbb{R}\!\)} &&&
       &{\footnotesize\(\!\omega_{\mathrm{pl}}^{\scriptscriptstyle\mathrm{(a)}}\)}
       &{\footnotesize\hspace{-0.3cm}\(\omega_{\mathrm{pl}}^{\scriptscriptstyle\mathrm{(b)}}\!\!\)}
\\\hline
	\(\epsilon_{\scriptscriptstyle\mathrm{Dru}}^{\phantom{|}}\)
	&\(\checkmark\) &\(\checkmark\) &\(\checkmark\) &\(\times\) &\(\checkmark\) &&
\\
	\(\epsilon_{\scriptscriptstyle\mathrm{Lin}}^{\mbox{\textcolor{Gray}{\tiny 2,3D}}}\)
	&\(\checkmark\) &\(\checkmark\) &\(\checkmark\) &\(\times\) &\(\checkmark\) &\(\times\) &\hspace{-0.5cm}\(\checkmark\)
\\
	\(\epsilon_{\scriptscriptstyle\mathrm{Me}}^{\mbox{\textcolor{Gray}{\tiny 2,3D}}}\)
	&\(\checkmark\) &\(\checkmark\) &\(\checkmark\) &\(\checkmark\) &\(\checkmark\) &\(\times\) &\hspace{-0.5cm}\(\times\)
\\
        \(\epsilon_{g_1}^{\mbox{\textcolor{Gray}{\tiny 2,3D}}}\)
	&\(\times\)     &\(\times\)     &\(\checkmark\) &\(\times\) &\(\times\) &\(\times\) &\hspace{-0.5cm}\(\times\) 
\\ 
        \(\epsilon_{g_{\mathrm{D}}}^{\mbox{\textcolor{Gray}{\tiny 2,3D}}}\)
	&\(\times\) &\(\checkmark\) &\(\checkmark\) &\(\times\) &\(\checkmark\) &\(\times\) &\hspace{-0.5cm}\(\times\)
\\
	\(\epsilon_{g_{E_{\mathrm{F}}}}^{\mbox{\tiny 2D}}\)
	&\(\times\)     &\(\checkmark\) &\(\times\) &\(\checkmark\) &\(\times\)     &\(\checkmark\) &\hspace{-0.5cm}\(\checkmark\) 
\\
	\(\epsilon_{g_{E_{\mathrm{F}}}}^{\mbox{\tiny 3D}}\)
	&\(\times\)     &\(\checkmark\) &\(\times\) &\(\checkmark\) &\(\times\)     &\(\times\) &\hspace{-0.5cm}\(\times\) 
\\
\end{tabular}
\renewcommand{\arraystretch}{1.0}
\caption{Exact criteria, satisfied (\(\!\checkmark\!\)) or not (\(\!\times\!\))
by the discussed approaches in an electron gas.
In Cols.1-4 (positive absorbance, real-valuedness on the imaginary \(\omega\) axis, causality, f-sum rule)
\(q\) is arbitrary.
Whether the \(q\!\to\!0\) plasmon coincides with the classical \(\omega_p\),
Col.\,5, differs in 2D and 3D.
}
\label{tbl: features}
\end{center}\end{table}

\goodbreak
Table~\ref{tbl: features} summarizes the behavior of (\ref{eq:chiLin}), (\ref{eq:Merm}), and (\ref{eq:chig}) with the choices listed above for \(g\).  Neither is fully satisfactory. 
As well-known \cite{giuliani2005quantum}, static and dynamic properties are not easily compatible.
Trying to extend the validity of Drude's model by replacing \(\omega_p^2/\omega^2\) with \(v\chi^0/(1\!-\!ig_{\scriptscriptstyle\mathrm{D}})\) 
fails due to the \(1/\omega\) divergence in \(g_{\scriptscriptstyle\mathrm{D}}(\omega\!\to\!0)\).
Although \(g_{\scriptscriptstyle\mathrm{E_F}}\) vanishes linearly for low \(\omega\), (\ref{eq:gEF}) has severe shortcomings where \(\mathrm{Re}\,\chi^0(q,\omega)\) changes sign.

\goodbreak
In case of a soft \(\omega_p(q\!\to\!0)\), as in the 2Deg, for ac\-cous\-tic plasmons, or in graphene, Eq.\,(\ref{eq:gEF}) ensures that the classical limit is reached \(\propto \eta^2\omega_p^2\).
We therefore study the 2Deg in further detail, and compare the above approaches for the sheet plasmon.

\section{Application to the 2Deg}
\label{sec: Application to the 2Deg}

The density \(n\) is related to the mean radius \(r^{\phantom{*}}_{\scriptscriptstyle\mathrm{S}} a_{\scriptscriptstyle\mathrm{B}}= 1/\sqrt{n\pi}\), with an effective Bohr radius \(a_{\mathrm{B}}\).
In typical semiconductor quantum wells \(r_{\mathrm{s}} \in[0.1,10]\).
Unless stated explicitly otherwise, we convert \(E_{\scriptscriptstyle\mathrm{F}}\) into eV for material parameters as 
given in Ref.\,\cite{Rugeramigabo08}
(a metallic \(\mathrm{DySi}_2\) monolayer on Si with  \(a^*_{\mathrm{B}}\!= 4.13 \,\mathrm{\AA}\),
\(n\!= 9\cdot\!10^{13}\)cm\(^{-2}\),
corresponding to \(E_{\scriptscriptstyle\mathrm{F}} \!= 269\,\)meV, Fermi wave vector \(k_{\scriptscriptstyle\mathrm{F}} \!= 0.23\,\mathrm{\AA}^{\!-1}\), and \(r_{\mathrm{s}}\!= 1.4\)).
An Ag monolayer also appears interesting \cite{SNDE11_lowenergyplasmons}.

\begin{figure}[t]
 \includegraphics[width=0.5\textwidth]{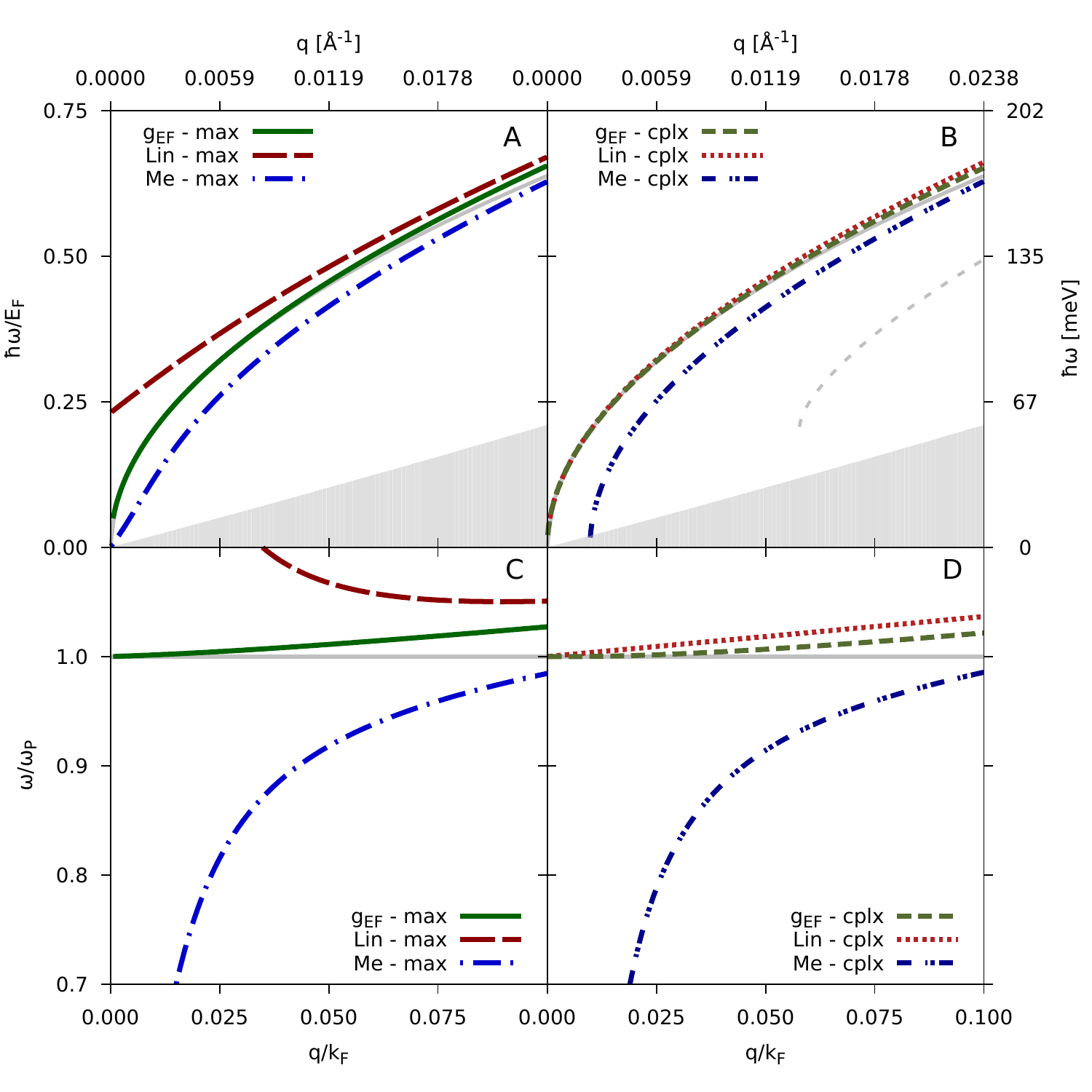}
\caption{
Plasmon dispersion \(\omega_{\mathrm{pl}}\) for \(r_{\mathrm{s}}\!=\!1.44\) and \(\hbar\eta\!=\!0.4\,E_{\scriptscriptstyle\mathrm{F}}\), 
obtained from Eqs.\,(1a) (`max') and (1b) (`cplx'). 
Mermin's results (blue, dash-[double]dotted) from (\protect{\ref{eq:Merm}})
are compared with those obtained from Eqs.\,(\protect{\ref{eq:chiLin}}) (dashed [A] and dotted [B] red) 
and (\protect{\ref{eq:gEF}}) ([full and dashed  green).
Light grey: Eq.\,(\ref{eq: plas_Reeps}) with (\ref{eq:Merm}).
Lower part: \(\omega_{\mathrm{pl}}\) divided by the classical 
\(\omega_p\propto\sqrt{q}\).
Upper part: \(\omega_{\mathrm{pl}}\) (both scales in \(E_{\scriptscriptstyle\mathrm F}\) and eV hold for both parts A,\,B); 
shaded area: e/h band (Landau damping).
}
\label{fig:plasmondispersionbrpa}
\end{figure}

In Fig.~\ref{fig:plasmondispersionbrpa} the plasmon dispersions are shown for a
fixed damping value of \(\eta\!= 0.4\,E_{\scriptscriptstyle\mathrm{F}}/\hbar
\approx 110\,\)meV/\(\hbar\).
We first compare the results from \(\epsilon_{\scriptscriptstyle\mathrm{Lin}}\) and \(\epsilon_{\scriptscriptstyle\mathrm{Me}}\).

For \(q\!\to\!0\), 
both, \(\omega_{\mathrm{pl}}^{\scriptscriptstyle\mathrm{(a)}}(q)\) and \(\omega_{\mathrm{pl}}^{\scriptscriptstyle\mathrm{(b)}}(q)\) as defined in Eqs.\,(\ref{eq: plas_Imchi})--(\ref{eq: plas_compl}), clearly deviate from the classical dispersion,
\(\omega_p= \sqrt{q\,ne^2/2m\varepsilon_{\scriptscriptstyle0}}\).
The Lindhard approach (\ref{eq:chiLin}) (red lines in Fig.\,\ref{fig:plasmondispersionbrpa}) yields 
an energy offset (long dashed line) via the maximum of the loss function, 
whereas \(\omega_p\) and \(\omega_{_\mathrm I}\) (dotted red line) from \(\epsilon\left(q,\,\omega_{_\mathrm I}\!+i\omega_{_\mathrm{II}}\right)=\!0\) almost coincide.

By contrast, the latter route gives a \(q-\)offset with Mermin's Eq.\,(\ref{eq:Merm}), (dash-double-dotted blue line),
where roots only occur beyond some finite \(q\) (as first noted in \cite{PhysRevB.29.2321}).
From the absorption maximum a \emph{linear} plasmon (dash--dotted blue line) follows, instead of \(\propto\sqrt{q}\).
The lower part of Fig.\,\ref{fig:plasmondispersionbrpa} makes this discrepancy evident.
To our knowledge, no such offsets have been yet observed.

The upper right part of Fig.\,\ref{fig:plasmondispersionbrpa} also shows the plasmon dispersion obtained from \(\mathrm{Re}\,\epsilon(q,\omega)=\!0\) (grey dashed line).
The result is depicted for \(\epsilon_{\scriptscriptstyle\mathrm{Me}}\).  Solutions exist only when both, \(\omega\) as well as \(q\), exceed some finite value. A big difference to the other curves is obvious.
Significant deviations also arise with \(\epsilon_{\mathrm{Lin}},\, \epsilon_{g_1},\, \epsilon_{g_\mathrm{D}},\,\)
small ones using \(\epsilon_{g_{E_\mathrm{F}}}\). 

The interpolation ansatz (\ref{eq:gEF}) (green curves) reproduces \(\omega_{p}\)
remarkably well via both routes (\emph{cf.} Fig.\,\ref{fig:plasmondispersionbrpa}\,C,\,D).
It thus appears to be the best choice. 

Finally, (\ref{eq: plas_magep}) 
yields a behavior 
(not displayed here)
similar to (\ref{eq: plas_Imchi}): 
\(\epsilon_{\scriptscriptstyle\mathrm{Lin}}\) giving an energy offset,
\(\epsilon_{\scriptscriptstyle\mathrm{Me}}\) leading to the vanishing of \(\omega^{\scriptscriptstyle\mathrm{(d)}}_{\scriptscriptstyle\mathrm{pl}}(q\!\to\!0)\) with a wrong order in \(q\),
and the results using \(g_{\scriptscriptstyle\mathrm{D}}\) being close to those from \(\epsilon_{\scriptscriptstyle\mathrm{Me}}\).
Again, \(g_{\scriptscriptstyle\mathrm{D}}\) obeys the classical limit, being a satisfying choice for the present purpose.

\begin{figure}[h!]
	\includegraphics[width=0.5\textwidth]{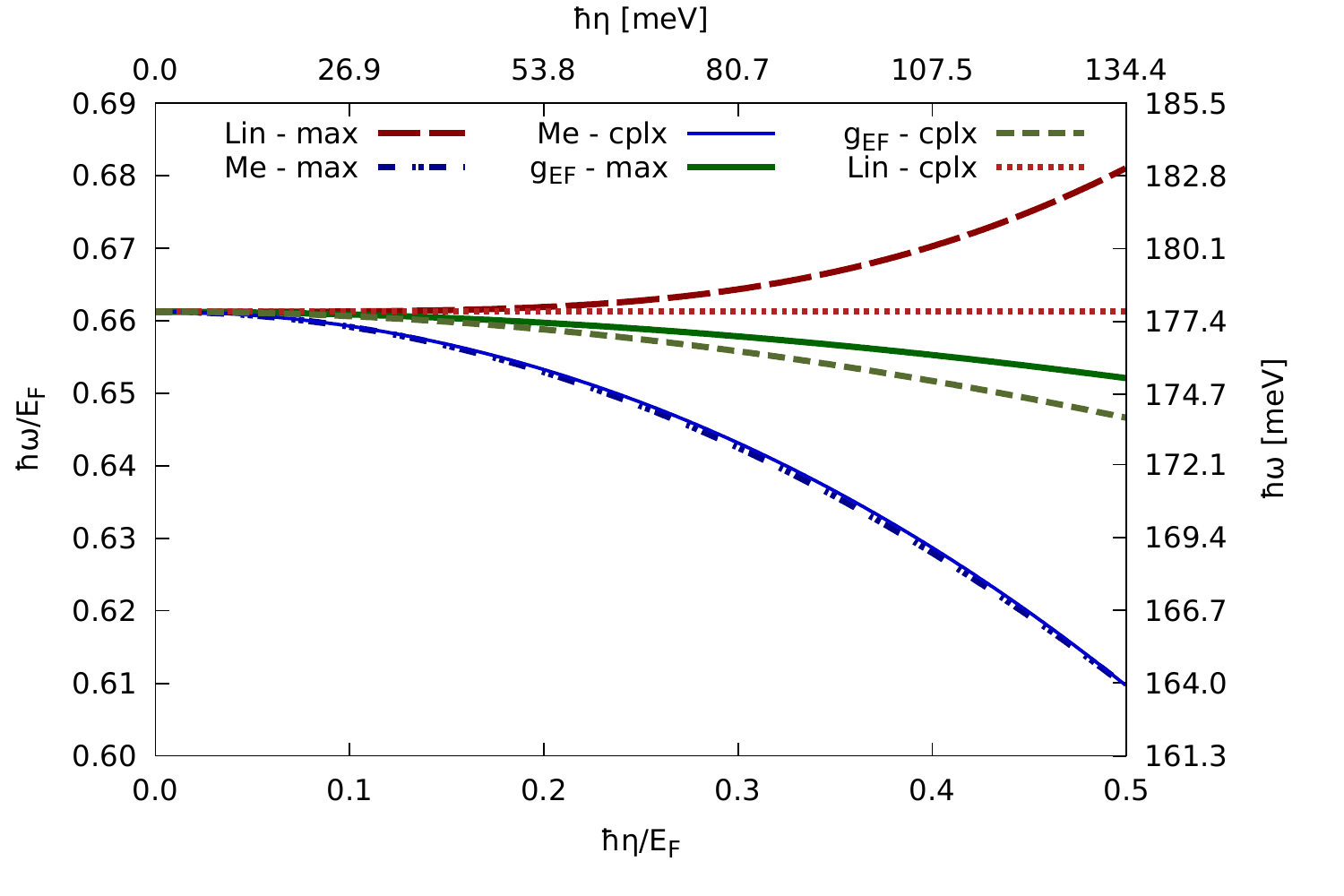}
\caption{
Plasmon dispersion \(\omega_{\mathrm{pl}}(q\!=\!0.1 k_{\scriptscriptstyle\mathrm{F}})\)
vs.\ inverse half width, at \(r_{\scriptscriptstyle\mathrm{S}}\!=1.44\).
Note the restricted \(\omega\) scale. Line styles are the same as in Fig.\,\protect{\ref{fig:plasmondispersionbrpa}}.
}
\label{fig:disperionoverbroad}
\end{figure}

Half widths as large as \(\sim\!500\)meV (\(\approx\!5\)\% of the peak position) were reported in \cite{nagaoplasmonL}.
In graphene \cite{LBPS10_pldampbelow}, even \(\sim\!50\)\% peak widths were found.
Figure~\ref{fig:disperionoverbroad} shows the predicted dispersions versus the damping parameter \(\eta\) for a fixed wave vector.
The difference between the results increases with the damping. 
Except for \(\epsilon_{\scriptscriptstyle\mathrm{Lin}}\) the dispersions decrease monotonically with \(\eta\).
A very close match holds for the \(\epsilon_{\mathrm{Me}}-\)plasmons based on criterion (\ref{eq: plas_Imchi}) or (\ref{eq: plas_compl}).

For \(\eta/\omega_{\mathrm{pl}}\!\approx\!10\)\% the differences are irrelevant,
for \(\eta/\omega_{\mathrm{pl}}\!\approx\!50\)\% the maximal distance amounts to \(\sim\!5\)\%.
In high resolution spectra of short-lived plasmons it therefore gets important, which expression is used for the plasmon.

The plain RPA is unreliable for higher \(q-\)values or dilute systems.
Much effort has gone into improvements over the decades \cite{giuliani2005quantum,Bhukal201513}, reviewing them is beyond our scope.
Commonly, a `local field corrected' or effective potential \(V(q)\!\equiv
v(q)\,(1\!-\!{\cal G}(q))\) is introduced in the dielectric function,
\begin{align}
    \epsilon_{_\mathrm{GRPA\!}} \>=\> 1- 
    \frac{v\,\chi^0}{1 + \big(1\!-\!{\cal{G}}\big)\,v\chi^0}
    \label{eq: GRPA} \;.
\end{align}
Choosing for \(V\) the particle--hole potential \cite{KroTrieste,KreilHDB15}
\begin{align}
    V_{\mathrm{ph}}(q) = \frac{\hbar^2q^2}{4mn}\Big(S(q)^{-2}-{S_{0}}(q)^{-2}\Big)\;,
    \label{eq: vphdef}
\end{align}
(\(S_0(q)\) is the non-interacting static structure factor \cite{giuliani2005quantum}), has the advantage of changing the sum rules (\ref{eq: w0SR}) and (\ref{eq: w1SR}) only marginally.
As \(V_{\mathrm{ph}}(q\!\to\!0)\) simplifies to \(v(q)\), the classical dispersion \(\omega_{p}\) is recovered for undamped systems.
A proper inclusion of correlation effects is ensured by using state-of-the-art ground--state  data for \(S(q)\) (\emph{e.g.} from \cite{gori2004pair,Asgari2004301}).

\begin{figure}
\includegraphics[width=0.5\textwidth]{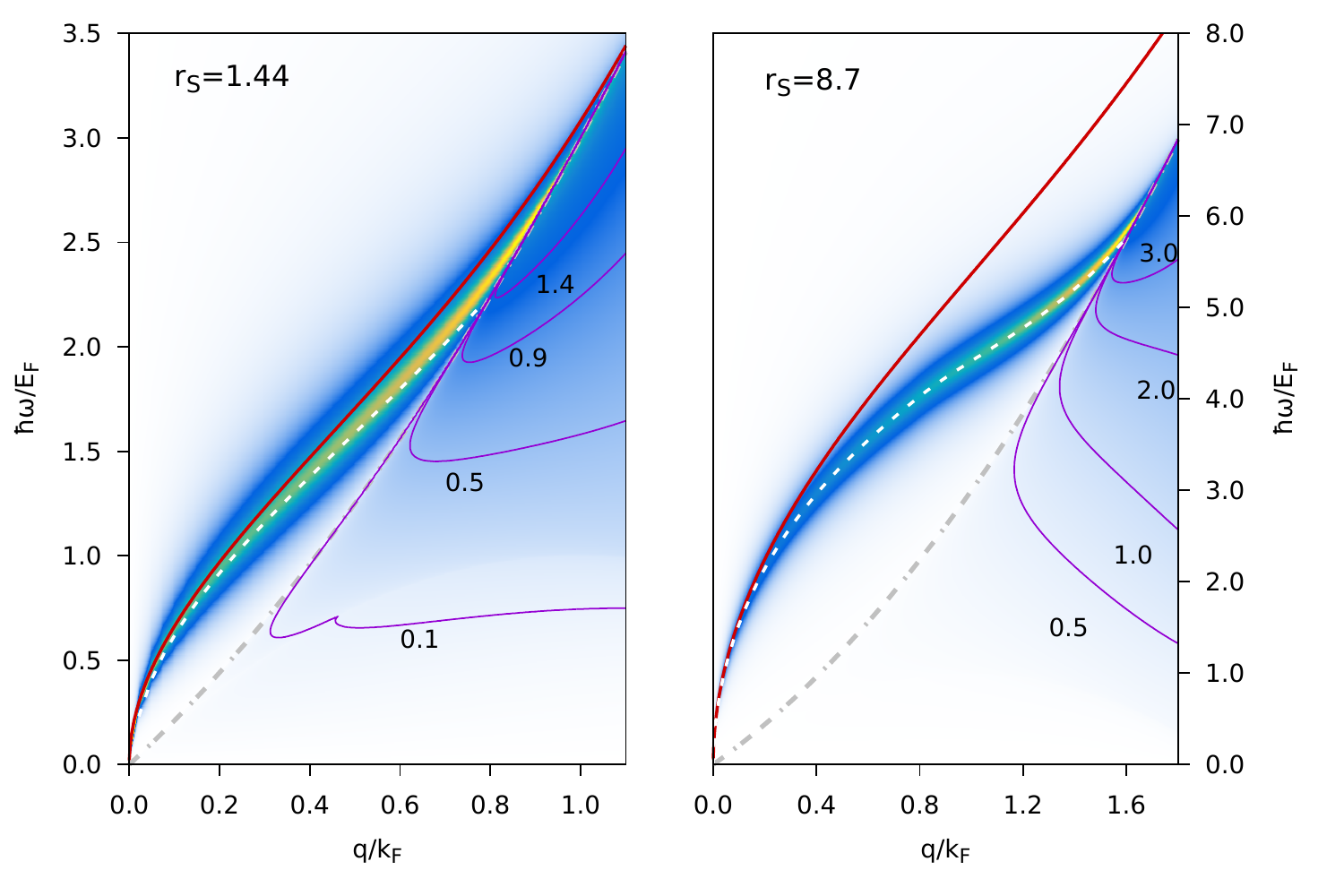}
\caption{
 Contour plot of \(-\mathrm{Im}\,\epsilon^{-1\!}_g(q,\omega)\) with (\ref{eq:gEF}) 
including screening (\ref{eq: vphdef}) with \(S(q)\) from \cite{gori2004pair}.
Densities are chosen as realized in \cite{Rugeramigabo08,hirjibehedin2002evidence}.
Red solid line: undamped bare RPA plasmon.
White dashed line: Mermin's collective mode from (\ref{eq:Merm}) with \(V_{\mathrm{ph}}\).
Grey dash--dotted line: upper e/h-band edge.
}
\label{fig:contourvphrs86}
\end{figure}

Correlations crucially lower the plasmon dispersion: in Fig.\,\ref{fig:contourvphrs86} the (yellow) maximum of the scattering loss function is significantly below the bare RPA result, Landau damping occurring at substantially lower \(q-\)vectors. 
The left part of Fig.\,\ref{fig:contourvphrs86} shows the same system as Fig.\,\ref{fig:plasmondispersionbrpa},
in the more dilute case (right part) the effect is even more pronounced.
There, \(\hbar\eta\!=\!0.4\,E_{\scriptscriptstyle\mathrm{F}}\!=\!0.05\,\)meV describes a relatively weaker damping (note the different \(\omega-\)scales).
The plasmon obtained from Eq.\,(\ref{eq: plas_Imchi}) using
\(\epsilon_{\scriptscriptstyle\mathrm{Me}}\) in the GRPA with \(V\) from
(\ref{eq: vphdef}) is found slightly lower (dashed line).

\section{Spin plasmon}
\label{sec: Spin plasmon}

Another mode can exist in electron layers: the spin--plasmon is the collective
excitation of the longitudinal magnetization, proportional to the spin density
\(s\!\equiv n_{\scriptscriptstyle\uparrow}\!-n_{\scriptscriptstyle\downarrow}\). 
Of negligible strength in \(\mathrm{Im}\,\epsilon^{-1}\), it manifests itself in the spin density response function \(\chi_{ss}\).
It may be observed \cite{agarwal2014long} in partially spin-polarized systems, as otherwise \(\chi_{ss}^{\scriptscriptstyle\mathrm{RPA}}=\chi^0\).
Its dispersion lies inside the e/h band of the majority spins \cite{KreilHDB15,agarwal2014long}.
Hence Landau--damping establishes a substantial natural broadening mechanism, we add no artificial broadening (\(\eta\!=\!0\)).
Again, the mode can be defined either via (a generalized) criterion (\ref{eq: plas_Imchi}) as a maximum in \(-\mathrm{Im}\,\chi_{ss}\),
or via criterion (\ref{eq: plas_compl}) as a zero of the denominator, in the RPA identical with \(\epsilon(q,\omega)\).
Such zeroes can be obtained graphically as intersections of the curves \(\epsilon_{\scriptscriptstyle\mathrm{I}}\!=\!0\) and \(\epsilon_{\scriptscriptstyle\mathrm{II}}\!=\!0\) in the complex \(\omega-\)plane.

\begin{figure}[h!]
\includegraphics[width=0.5\textwidth]{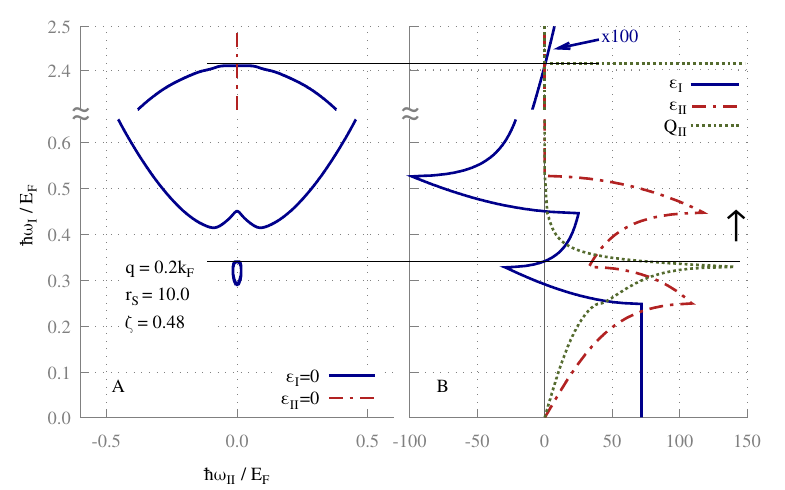}
\caption{Spin--plasmon determination in a 2Deg with \(\lesssim75\%\) spin-\(\scriptstyle\uparrow\) electrons.
A: roots of \(\epsilon_{\scriptscriptstyle\mathrm{I}}\) (full blue line) and \(\epsilon_{\scriptscriptstyle\mathrm{II}}\) (dash--dotted red line) in the \(\omega_{\scriptscriptstyle\mathrm{I}}\!+i\omega_{\scriptscriptstyle\mathrm{II}}\) plane.
B: Shaded regions with \(\scriptstyle\uparrow\,\)(\(\scriptstyle\downarrow\)) arrows indicate the e/h bands for spin-\(\scriptstyle\uparrow\) (-\(\scriptstyle\downarrow\)) particles.
Full blue and dash--dotted red line: \(\mathrm{Re}\) and \(\mathrm{Im}\) of \(\epsilon(0.2k_{\scriptscriptstyle\mathrm{F}},\omega)\), respectively.
Horizontal lines: density- and spin-plasmon from \(\epsilon_{\scriptscriptstyle\mathrm{I}}\!=\!0\). 
Green dotted line: \(-\mathrm{Im}\,v\,\chi_{ss}\).
}
\label{fig:spincomparision}
\end{figure}

This is seen in Fig.\,\ref{fig:spincomparision}.A for a 2Deg with a density of \(r_{\scriptscriptstyle\mathrm{S}}\!=10\) and spin imbalance \(s\!= 0.48\,n \equiv n\zeta\). 
For the conven\-tional (charge-)density-plasmon the intersection is clearly visible (due to \(\eta\!=\!0\) found at \(\omega_{\scriptscriptstyle\mathrm{II}}\!=\!0\)). 
The right part, Fig.\,\ref{fig:spincomparision}.B, shows
\(\epsilon_{\scriptscriptstyle\mathrm{I}}\) and
\(\epsilon_{\scriptscriptstyle\mathrm{II}}\) vs.\ real \(\omega\) (on the
vertical axis), the top horizontal line marks the plasmon position .
However, there is no (further) vanishing of \(\epsilon_{\scriptscriptstyle\mathrm{II}}\) within the majority e/h excitation band (light grey region).
Though not a `true' collective mode, in \(-v(q)\,\mathrm{Im}\,\chi_{ss}(q,\omega)\) (dotted green line) the spin plasmon is obvious as a sharp maximum \emph{outside} the minority band; distinctly below the zero of \(\epsilon_{\scriptscriptstyle\mathrm{I}}\).
Again, Eqs.\,(\ref{eq: plas_Imchi},\ref{eq: plas_Reeps}) yield noticeably different values.

\section{Conclusion}

We gave a brief overview on the formal properties of various response functions phenomenologically accounting for plasmon damping.
Many common approaches do not exactly recover the classical limit.
This can be ensured via Eqs.\,(\ref{eq:gEF}) and (\ref{eq:chig}) in the 2Deg, where we quantitatively studied the influence of the damping on the dispersion.
Special emphasis lay on the distinct results from different criteria; albeit
small, this gets relevant with state-of-the-art measurements having reached the experimental resolution.
Similar conclusions can be drawn for local--field corrected dielectric functions and for the spin-plasmon, 
and are expected to also hold for graphene.

\section*{Acknowledgment}
We thank Nikita Arnold for helpful discussions and the W.\ Macke Stipendienstiftung for financial support.

\section*{References}

\bibliographystyle{model1a-num-names}
\bibliography{references}

\end{document}